\documentclass[sigconf,table]{acmart}

\usepackage{booktabs} % For formal tables
\usepackage{mathtools}
\usepackage{hhline}
\usepackage{pbox}
\usepackage{multirow}
\usepackage{slashbox}
\usepackage{bm}
\usepackage{algorithmicx}
\usepackage{algorithm}
\usepackage{algpseudocode}
\usepackage{balance}

\newcolumntype{C}[1]{>{\centering}m{#1}}

\begin{document}
\title{Learning from Online Regrets: From Deleted Posts to Risk Awareness in Social Network Sites}

\author{Nicol\'{a}s Emilio D\'{i}az Ferreyra}
\affiliation{University of Duisburg-Essen, Germany\\RTG User-Centred Social Media\\https://www.ucsm.info/\\
}
\email{nicolas.diaz-ferreyra@uni-due.de}

\author{Rene Meis}
\affiliation{University of Duisburg-Essen, Germany\\RTG User-Centred Social Media\\https://www.ucsm.info/\\
}
\email{rene.meis@uni-due.de}

\author{Maritta Heisel}
\affiliation{University of Duisburg-Essen, Germany\\RTG User-Centred Social Media\\https://www.ucsm.info/\\
}
\email{maritta.heisel@uni-due.de}

% The default list of authors is too long for headers}

\renewcommand{\shortauthors}{N. E. D\'{i}az Ferreyra et al.}
\renewcommand{\shorttitle}{Learning from Online Regrets: From Deleted Posts to Risk Awareness in SNSs}

\begin{abstract}
Social Network Sites (SNSs) like Facebook or Instagram are spaces where people expose their lives to wide and diverse audiences. This practice can lead to unwanted incidents such as reputation damage, job loss or harassment when pieces of private information reach unintended recipients. As a consequence, users often regret to have posted private information in these platforms and proceed to delete such content after having a negative experience. Risk awareness is a strategy that can be used to persuade users towards safer privacy decisions. However, many risk awareness technologies for SNSs assume that information about risks is retrieved and measured by an expert in the field. Consequently, risk estimation is an activity that is often passed over despite its importance. In this work we introduce an approach that employs deleted posts as risk information vehicles to measure the \textit{frequency} and \textit{consequence} level of self-disclosure patterns in SNSs. In this method, consequence is reported by the users through an ordinal scale and used later on to compute a risk criticality index. We thereupon show how this index can serve in the design of adaptive privacy nudges for SNSs. %serve in the generation of adaptive risk warnings.
\end{abstract}

%
% The code below should be generated by the tool at
% http://dl.acm.org/ccs.cfm
% Please copy and paste the code instead of the example below. 
%



\copyrightyear{2019} 
\acmYear{2019} 
\setcopyright{acmcopyright}
\acmConference[UMAP'19 Adjunct]{27th Conference on User Modeling, Adaptation and Personalization Adjunct}{June 9--12, 2019}{Larnaca, Cyprus}
\acmBooktitle{27th Conference on User Modeling, Adaptation and Personalization Adjunct (UMAP'19 Adjunct), June 9--12, 2019, Larnaca, Cyprus}
\acmPrice{15.00}
\acmDOI{10.1145/3314183.3323849}
\acmISBN{978-1-4503-6711-0/19/06}

\maketitle

\section{\uppercase{Introduction}} \label{introduction}

\noindent Social Network Sites (SNSs) like Facebook or Twitter allow users to create and maintain social connections with a wide spectrum of online communities which represent (in many cases) the different facets of their lives. User-generated content plays a major role in this process since posts, comments, videos and photos are the vehicles that allow users to relate with each other within these platforms. Nevertheless, disclosing private and sensitive information through these communication channels can result in unwanted incidents such as reputation damage, unjustified discrimination or even job loss when such content reaches an unintended audience \cite{wang2011regretted,christofides2012risky}. Consequently, users very often regret having posted private information in SNSs because they were unable to anticipate the negative consequences of their actions \cite{wang2011regretted}.

Risk awareness is key for making better and more informed decisions in our daily lives. For instance, being aware of the risks of smoking can discourage people in engaging with tobacco consumption \cite{hiilamo2014evolution}. Likewise, nutrition labels can support people in improving their eating habits \cite{downs2009strategies}. However, users of SNSs receive very little (for not saying none) information about the risk of online interaction, neither as part of the platform's layout nor in the body of the privacy policy \cite{diazferreyra2017should}. Moreover, SNSs very often present themselves as spheres free of intrusions and privacy risks. This lack of information modulates the perceived severity of privacy risks in favour of information disclosure and, consequently, in benefit of the service providers \cite{stark2016emotional,samat2017formatvscontent}.

%to induce behavioral changes among users of SNSs when making privacy-related decisions
%Aditionaly, interventions generated by PTs can be more efective when tailored according to the privacy goals of each user.
Privacy scholars have developed a wide variety of \textit{preventative technologies} \cite{ziegeldorf2015comparison,wizards2010fang,diazferreyra2017online} to induce changes in the privacy decisions made by the users of SNSs. For instance, these technologies provide cues about the semantics of the content being shared by the users (i.e. if a post contains private information or not) in order to persuade them towards safer privacy practices \cite{diazferreyra2017should}. Such approach can be improved by personalizing these cues with risk information associated with self-disclosure patterns in SNSs \cite{diazferreyra2017online,schaewel2018disclose}. That is, providing a personalised assessment of the risks that may take place if private information is revealed in a post. A prerequisite to perform this task is to have a repository of unwanted incidents, together with their respective frequencies and severity levels \cite{diaz2018iwse}. However -to the best of our knowledge-, not many efforts have been made on defining the necessary mechanisms to collect and process such information. In other words, the information necessary to generate a proper risk estimation often remains as an assumption for preventative technologies.

%Such approach can be improved by communicating  well-known privacy risks that are associated with patterns of information disclosure in SNSs

In this work, we introduce a method to estimate the risks of information disclosure in SNSs using deleted posts as indicators of online regrets. Since users often delete their publications after living a negative experience in SNSs \cite{wang2011regretted}, we propose to leverage such deleted content to (i) retrieve information about unwanted incidents, and (ii) estimate their respective frequency and consequence level. For this purpose, we introduce an interface for collecting this information in which the perceived consequence level of the incidents can be entered by the users using nominal values (i.e.  \textit{catastrophic}, \textit{major}, \textit{moderate}, \textit{minor} and \textit{insignificant}). In line with this, we describe how this information can be used later on to compute a risk criticality index and integrate it into an adaptive awareness mechanism.

The rest of the paper is organized as follows. In the next section we discuss related work in the area of adaptive privacy awareness. Following, section \ref{background} introduces the theoretical foundations of this paper. In particular, we discuss the use of heuristics for the generation of awareness together with a risk estimation approach. Section \ref{method} elaborates on a method for collecting evidence of recurrent unwanted incidents in SNSs using deleted posts as risk-information vehicles. Next, section \ref{algorithm} introduces an algorithm for the generation of adaptive privacy awareness. This algorithm combines the output produced by the method of section \ref{method} together with the risk index introduced in section \ref{background}. Following, section \ref{discussion} analyses the strengths and limitations of our approach, and section \ref{evaluation} describes the corresponding design and evaluation plan. Finally, in section \ref{conclusion} we outline the conclusions of this paper and give directions for future work.

\section{\uppercase{Related Work}}

\noindent Privacy in SNSs is a multifaceted issue that has caught the attention of many researchers across different disciplines \cite{diaz2012understanding}. In the particular case of online self-disclosure, several  Preventative Technologies (PTs) have been proposed for nudging the users towards a safer privacy behavior \cite{diazferreyra2017should}. Basically, PTs generate \textit{interventions} (i.e. warning messages or suggestions) when users attempt to publish private or sensitive information in their profiles \cite{diazferreyra2017should}. This way, users are induced to reflect on the content they are about to share and the negative consequences that may occur after posting such content. Although this is a well-grounded persuasive strategy, warnings are sometimes perceived as too invasive or annoying by the users. This happens basically because not all users have the same privacy attitudes or concerns and, consequently, adopt different privacy strategies \cite{schaewel2017paving}. For instance, some users are more willing to disclose private information without much concern about the consequences, and others rather keep such information away from unwanted recipients. This suggests that PTs should generate interventions aligned with the users' privacy attitudes in order to engage them in a continuous learning process \cite{diaz2018iwse}. In other words, PTs should incorporate \textit{adaptivity} principles into their design.

\begin{figure}[!h]
\centering
\includegraphics[height=3.5cm]{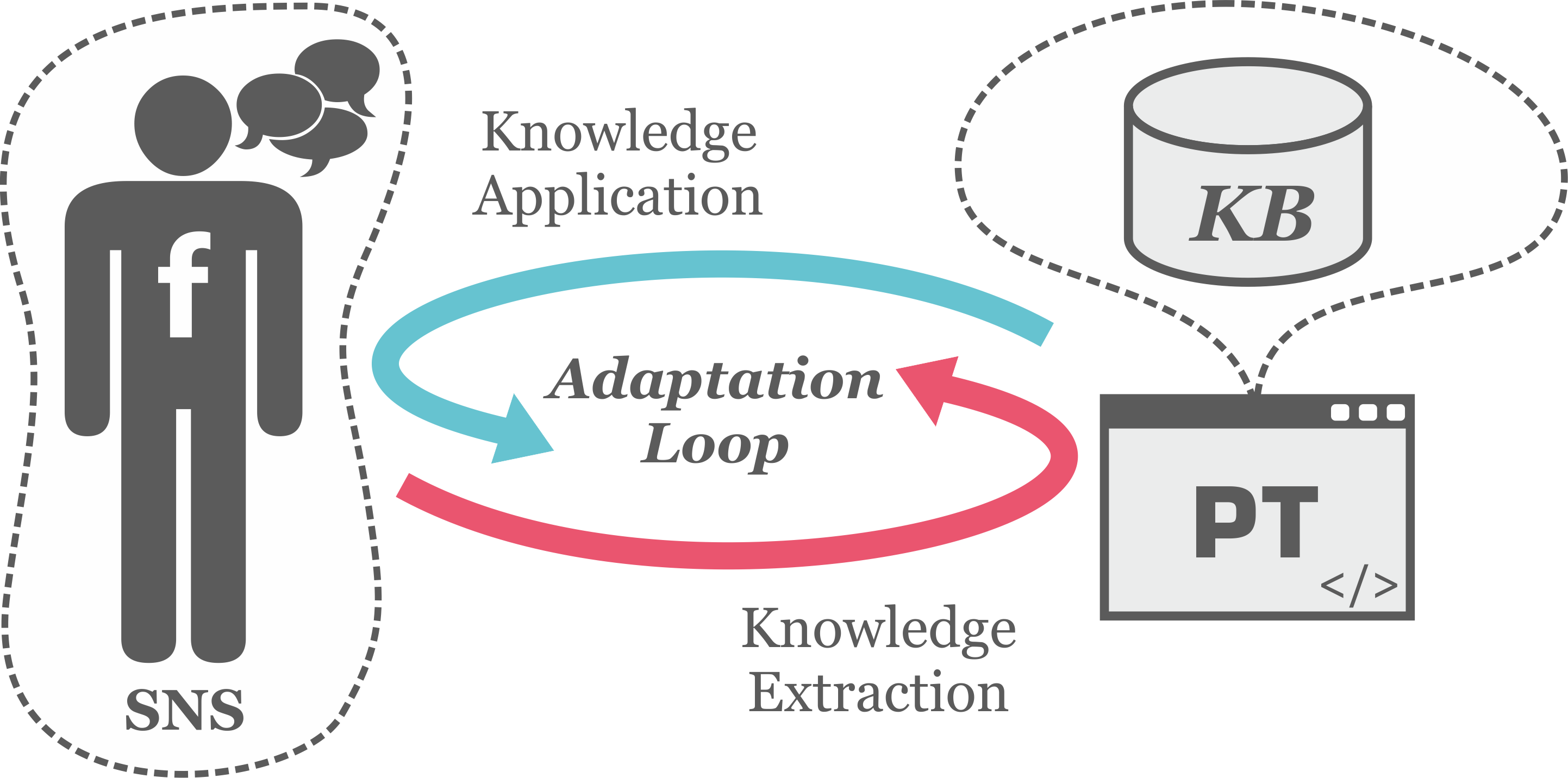}
\caption{Adaptive Privacy Loop}
\label{fig:1}
\end{figure}

At a glance, the adaptive awareness process of PTs can be described as a loop consisting of two main activities as shown in Fig.~\ref{fig:1}. In the first step, \textit{knowledge extraction}, the information that is necessary for the generation of adaptive interventions is gathered and stored inside a Knowledge Base (KB). The second step, \textit{knowledge application}, consists of querying the information inside the KB to shape personalized interventions. This two-phase process repeats itself after some time in order to update the information inside the KB and thereby improve the quality of the interventions. For instance, Misra et al. \cite{misra2016non} developed an approach in which the emerging communities inside a user's ego network (i.e. the network of connections between his/her friends) is used to build personalized access-control lists (i.e. \textit{black lists} of information recipients). In this case, knowledge extraction consists of retrieving the user's ego-network, and knowledge application involves the use of this information for generating a personalized access-control list. This process repeats itself when the topology of the user's ego-network changes (i.e. when contacts or links between contacts are added/removed).

Combining adaptive awareness with risk management features is a promising approach for promoting safer privacy decisions among the users of SNSs. As mentioned by D\'{i}az Ferreyra et al. \cite{diaz2018iwse} and Acquisti et al. \cite{samat2017formatvscontent}, information about the potential risks of a self-disclosure act can nudge the users towards more proactive privacy decisions. In line with this premise (i.e. more risk information, better privacy decisions), De et al. \cite{deMetayer2018iwpe} developed an approach using attack-trees to inform the users of SNSs about the privacy risks that may result from their privacy settings (e.g. the risks of having a public profile). Despite its novelty, assumptions were made with regard to the information used for the estimation of such risks. This is, a prerequisite for the application of this approach is to have a KB containing information about common unwanted incidents, their frequency and consequence levels. In order to endow this and other PTs with the information necessary for computing privacy risks, we propose a method for collecting such risk-related information using deleted posts. Likewise, we describe how this information can be aggregated and used to define an adaptive mechanism of risk awareness.

\section{\uppercase{Background}} \label{background}

\noindent In this section we introduce the theoretical foundations of this paper. Particularly, we discuss how regrettable online experiences can be translated into patterns of information disclosure and used thereafter for the generation of privacy awareness inside SNSs. In line with this, we discuss the importance that the estimation of risk values has for PTs and introduce an approach that can be used to carry on with this task.

\subsection{Privacy Heuristics} \label{heuristics}

\noindent A privacy regret in SNSs can be defined as a \textit{``feeling of sadness, repentance or disappointment which occurs when a piece of sensitive information reaches an unintended audience and results in an unwanted incident''} \cite{diaz2018iwse}. For instance, the regrettable scenario of Fig. \ref{fig:fig2} illustrates a situation in which a user gets in trouble with her employer after posting a negative comment about her workplace. Probably, this scenario might have been experienced by more than one user while interacting inside a SNS. In this case, such scenario can be abstracted into a pattern of information disclosure that represents this and other regrettable scenarios with similar characteristics. Diaz Ferreyra et al. \cite{diaz2018iwse} propose to describe recurrent self-disclosure scenarios using \textit{privacy heuristics} (PHs). Basically, PHs model patterns of information disclosure as a tuple \textit{\textless PAs, Audience, Risk\textgreater} where \textit{PAs} is a set of private attributes, \textit{Audience} is a collection of recipients (e.g. Facebook friends), and \textit{Risk} is a characterization of the severity (i.e. a measure of the \textit{consequence} and \textit{frequency}) of an \textit{Unwanted Incident} (UIN). Hence, the corresponding PH for the scenario of Fig. \ref{fig:fig2} models the severity of job loss when a negative comment about ones' workplace is disclosed to an audience composed by work colleagues.

The knowledge inside PHs can be used by PTs to communicate the risks associated with disclosing certain patterns of private information in SNSs. This can be done by checking the \textit{Risk} information inside a PH whose \textit{PAs} match with the ones disclosed in a new post \cite{diaz2018iwse}. For instance, if the same user of Fig. \ref{fig:fig2} (or any other user) attempts to disclose a negative comment about her workplace inside a new post, the PT could inform her that this can lead to job loss if seen by her work colleagues. This information flow corresponds to the knowledge application step described in Fig. \ref{fig:1} for a PT whose KB consists of a collection of PHs. Such a Privacy Heuristics Data Base (PHDB) can be engineered through the elicitation of regrettable scenarios and their later encoding into PHs \cite{diaz2018iwse,diazferreyra2017online}. In principle, one could shape a PH out of the experience of a single user (e.g. extracting the PAs, the Audience and the Risks in a face-to face interview or through an online questionnaire) \cite{diazferreyra2017online}. However, the consequence level of an UIN is subjective (i.e. varies from individual to individual) and the same UIN can be perceived as insignificant by one user and catastrophic by others. Moreover, a single occurrence of an UIN is not enough to estimate its frequency. Therefore, one must have multiple sources of evidence of an UIN to generate a good risk estimator.

\begin{figure}[h]
\def\arraystretch{0.7}
\setlength{\fboxsep}{8pt}%
\fbox{%
	\begin{minipage}{7.8cm}
	\begin{center}
	\textbf{USER'S POST} 
	\end{center} \textit{``A typical day at the office. Lots of complaints and bad mood. Cannot wait for the day to be over...!''}
		\rule{7.8cm}{0.4pt}{\vspace{2mm}}
		\textbf{Actual Audience:} PUBLIC.\\
		\textbf{Unintended Audience:} The user's work colleagues.\\
		\textbf{Unwanted Incidents:} Reputation damage; job loss.
	\end{minipage}}
\caption{ Example of self-disclosure scenario}
\label{fig:fig2}
\end{figure}

\subsection{Criticality Index} \label{critIndex}

\noindent Estimating the severity of privacy risks is an important step towards the generation of privacy awareness. Basically, this allows to prioritize the communication of those risks with a high severity level over those risks whose severity level is low. One way to estimate such risk levels is through a \textit{risk index} that aggregates instances of elementary risk evidence measured through quantitative or qualitative data \cite{facchinetti2018risk}. Since ordinal scales such as \textit{insignificant}, \textit{minor}, \textit{moderate}, \textit{major} and \textit{catastrophic} are convenient when measuring the consequence of unwanted incidents \cite{lund2010model}, it is desirable that a risk index can deal with ordinal variables. One approach that takes this aspect into account is the \textit{Criticality Index} (CI) introduced by Facchinetti et al. \cite{facchinetti2018risk} which generates a normalized value $I$ of risk taking as input the frequency of the values used to measure the consequence of an unwanted incident. This is, given a categorical random variable $X$ with ordered categories $x_k$ that represent decreasing consequence levels $k=1,2,...,K$, a value of $I$ closer to 0 indicates that the severity of a risk event is likely to be low whereas values closer to 1 indicate that the severity is likely to be high. An \textit{estimator} of the risk index $I$ can be obtained out of a sample of size $n$ of the categorical variable $X$ with the following equation \cite{facchinetti2018risk}: %\cite{facchinetti2018risk,facchinetti2018cybersecurity} - \cite{facchinetti2018cybersecurity}

\begin{gather}
\hat{I}=\frac{\sum_{k=1}^{K}\tilde{F_k}-1}{K-1} \tag{1}
\end{gather} %\shortintertext{where }

where, for a consequence scale of $K$ levels, the values $k=1$ and $k=K$ correspond to the highest and lowest consequence values of an unwanted incident, respectively. Likewise, $\tilde{F_k}$ corresponds to the \textit{empirical} distribution function of the random variable $X$, which for a category $x_k$ is computed as the number of observations $r_l$ in the sample with consequence levels between $1$ and $k$:

\begin{gather*}
\tilde{F_k} = \sum\nolimits_{l=1}^{k} \frac{r_l}{n} \quad \textrm{for} \quad  k=1,2,...,K
\end{gather*}

Eq. 1 aggregates evidence about the consequence of an unwanted incident to determine the severity of its corresponding risk event. Consequently, $\hat{I}$ can be used to instantiate the Risk component of a PH, and consequently to generate risk awareness on well-known patterns of information disclosure in SNSs. Under this premise, we will describe in the next two sections (i) a method for collecting evidence of unwanted incidents in SNSs using deleted posts and (ii) instantiate the Risks of PHs for the later generation of adaptive risk awareness in SNSs. This second task, which involves the generation of a confidence interval that contains the real value of $I$, is addressed in Section \ref{algorithm}.

\section{Learning from Deleted Posts} \label{method}

%As we discussed in Section \ref{background}, PHs are promising instruments for the generation of risk awareness in SNSs. However, in order to put these instruments into practice, one must count with a good estimation about the severity of privacy risks associated with well-known patterns of information disclosure. In this section we introduce a method for collecting information about regrettable self-disclosure experiences in SNSs and use it to estimate the corresponding risk criticality index. Such index, is used to instantiate the Risk component of PHs and thereafter in the generation of adaptive risk awareness in SNSs.

As we discussed in Section \ref{background}, PHs are promising instruments for the generation of risk awareness in SNSs. However, in order to put these instruments into practice, one must properly estimate the severity of the privacy risks that are associated to them. The CI discussed in section \ref{critIndex} is an adequate instrument to perform such an estimation; however, this requires empirical evidence about the frequency of UINs. In order to gather the evidence necessary to estimate those privacy risks that are associated with PHs we have defined a method consisting of five steps: \textit{Analyse Post Content}, \textit{Elicit Unwanted Incident}, \textit{Match Existing Heuristics}, \textit{Add New Heuristic}, and \textit{Update Contingency Table}. As depicted in Fig.~\ref{fig:fig4}, each stage of the method draws on different external inputs and generates the outputs for the next step. The final output of the method is an updated version of a \textit{contingency table} which is a data structure used to summarize the frequency of the different UINs that are associated with a PH.

%\subsection{The Method} \label{method}

%In order to estimate those privacy risks that are associated with PHs we have defined a method consisting of five steps: \textit{Analyse Post Content}, \textit{Elicit Unwanted Incident}, \textit{Match Existing Heuristics}, \textit{Add New Heuristic}, and \textit{Update Contingency Table}. As depicted in Fig.~\ref{fig:fig4}, each stage of the method draws on different external inputs and generates the outputs for the next step. The final output of the method is an updated version of a \textit{contingency table} which is a data structure used to summarize the frequency of the different UINs that are associated with a PH.

%The final output of the method is an updated version of a \textit{contingency table} (CT) which is a data structure used to summarize the severity of different privacy risks that are associated with PHs.

\textbf{Step 1: Content Analysis} The method starts when a post with private information is deleted by the user. Basically, an event of such characteristics is likely to occur when the user has lived a regrettable experience after disclosing sensitive data to the wrong audience. Therefore, a deleted post which encloses this type of information can be used as a vehicle for gathering information about UINs that result from self-disclosure actions in SNSs. To start with the identification of private information inside deleted posts, one must have a taxonomy of attributes of sensitive nature. This is often a challenge itself due to the multiple definitions of private information, and the influence that the context where such information is disclosed can have for this type of analysis. Consequently, different taxonomies of private/sensitive attributes have been proposed by scholars, each of them based on different interpretations of private information \cite{petkos2015pscore}. Since the goal of this step is to identify regrettable posts, the taxonomy of attributes used for this task should be aligned with this purpose. Thus, such taxonomy must include attributes for which there is evidence of regret when they have been disclosed inside SNSs. 

\begin{table}[!b]
\centering
\bgroup
\def\arraystretch{1.1}
\resizebox{\linewidth}{!}{
\begin{tabular}{ |p{2.3cm}|p{5.5cm}|  }
 \hline
% \multicolumn{3}{|c|}{Self-disclosure Dimensions} \\
% \hline
 \textbf{Dimension}& \textbf{Surveillance Attributes}\\
 \hline
 Demographics&  Age, Gender, Nationality, Racial origin, Ethnicity, Literacy level, Employment status, Income level, Family status \\
 \hline
 Sexual Profile& Sexual preference \\
 \hline
 Political Attitudes & Supported party, Political ideology \\
 \hline
 Religious Beliefs & Supported religion \\
 \hline
 Health Factors and Condition&  Smoking, Alcohol drinking, Drug use, Chronic diseases, Disabilities, Other health factors  \\
 \hline
 Location&  Home location, Work location, Favorite places, Visited places \\
 \hline
 Administrative& Personal Identification Number \\
  \hline
 Contact& Email address, Phone number \\
 \hline
 Sentiment& Negative, Neutral, Positive \\
 \hline
\end{tabular}}
\egroup
\bigskip
\caption{The ``self-disclosure'' dimensions.}
\label{table:1}
\end{table}

Based on a study of regrets conducted by Wang et al. \cite{wang2011regretted}, D\'{i}az Ferreyra et al. \cite{diazferreyra2017online} proposed a taxonomy of \textit{surveillance attributes} (SAs) that provides an intuitive representation of different aspects of the users' private information in SNS. As shown in Table \ref{table:1}, this taxonomy organizes a collection of personal attributes (i.e. the SAs) around a number of high-level categories called ``self-disclosure dimensions'' which arrange them into \textit{demographics, sexual profile, political attitudes, religious beliefs, health factors and condition, location, administrative, contact, and sentiment}. We will adopt this taxonomy for the identification of private information, meaning that this step can be summarized as the identification of SAs inside deleted posts. Taking the example of Fig \ref{fig:fig2}, the SAs disclosed in the post are \textit{work location}, \textit{employment status} together with a \textit{negative sentiment}. In principle, these and the rest of the taxonomy's SAs could be automatically identified using methods and techniques for Natural Language Processing (NLP) such as regular expressions, named-entity recognition or deep learning algorithms (e.g. Nguyen-Son et al. \cite{nguyen2015anonymizing} applied support vector machines for the identification of private information in SNSs). This task goes beyond the scope of this paper; therefore, we rely on the assumption that this step can be automated.

\begin{figure}[!h]
\centering
\includegraphics[height=3.6cm]{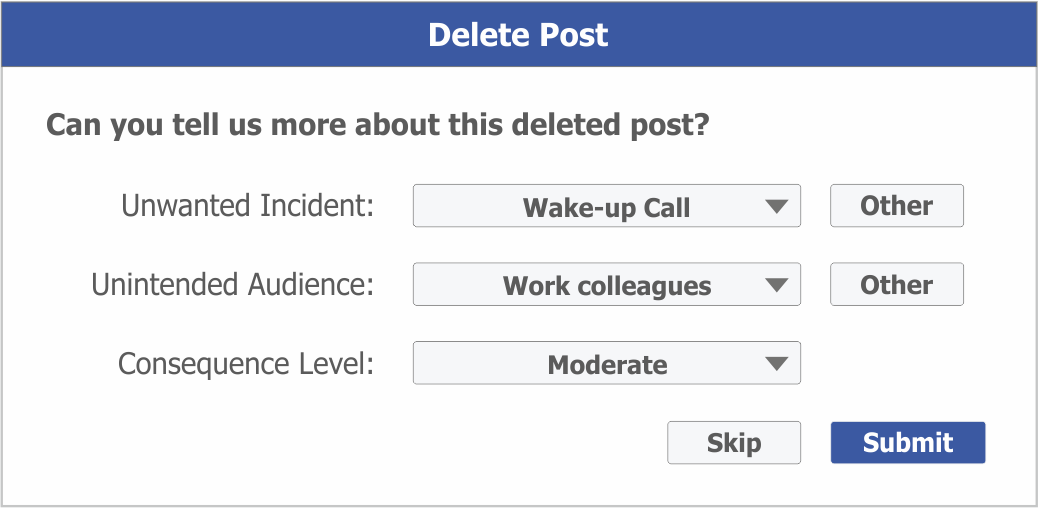}
\caption{Submit Unwanted Incidents Interface}
\label{fig:3}
\end{figure}

\begin{figure*}[t]
\centering
\includegraphics[height=4.85cm]{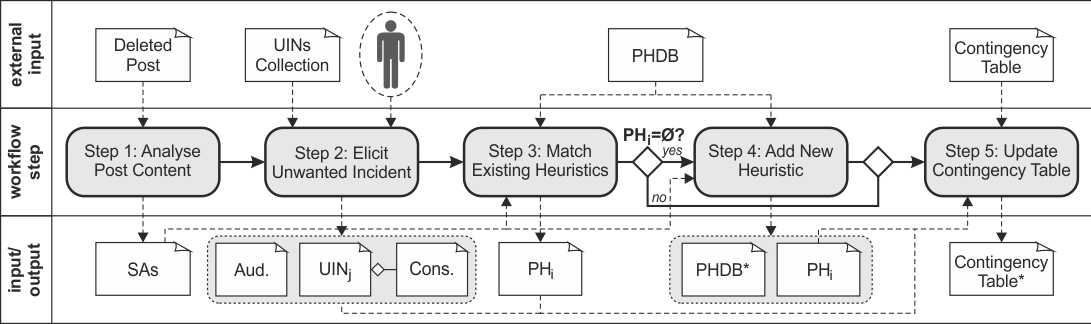}
\caption{Update Contingency Table Method}
\label{fig:fig4}
\end{figure*}

\textbf{Step 2: Elicit Unwanted Incident} After identifying a set of SAs inside the deleted post the next step is to gather information about the reasons that led the user to delete such content. In other words, we must (i) confirm that the post was deleted by the user because its publication led to an UIN, and (ii) if true, determine which particular UIN took place together with its consequence level. For this purpose, we have designed the interface of Fig \ref{fig:3}, which is displayed to the user after she deletes a post which contains SAs. Basically, this interface asks the user if the post being deleted caused her an unpleasant experience and asks her to provide further details. Particularly, the user can specify the unwanted incident that took place using a pre-defined list of UINs (i.e. using the list-box with the label ``Unwanted Incident'') or adding a new item if it is not in the list (i.e. using the ``Other'' button). Likewise, the user can select the audience that should have not seen the post using a pre-defined list of social circles (i.e. using the list-box with the label ``Unintended Audience'') or adding a new item if it is not in the list (i.e. using the ``Other'' button). Finally, the consequence level of the unwanted incident can be specified using the list-box with the label ``Consequence Level''. Fig. \ref{fig:3} illustrates the report corresponding to the scenario of Fig. \ref{fig:fig2} in which the user describes the wake-up call from her superior as an UIN with a \textit{moderate} consequence level. After completing all the fields on the interface, the user can submit the report on the deleted post by clicking on ``Submit''.

\textbf{Step 3: Match Existing Heuristics} As mentioned in section \ref{heuristics}, a PHDB is a KB that contains a collection of PHs which represent different regrettable self-disclosure scenarios. Basically, these scenarios are patterns that repeat themselves when different posts with similar information result in the same UIN after reaching the same social circle (i.e. the same Audience). For instance, a user who posts \textit{``I hate my job at this company but damn, it pays the rent! \#keepcalm''} is revealing the same set of SAs like the ones being revealed in the post of Fig.~\ref{fig:fig2}. Moreover, this user can suffer the same UIN as the user of Fig.~\ref{fig:fig2} if the post is seen by her colleagues from work. Consequently, the same pattern (i.e. the same PH) can be extracted from two or more different posts. When this happens, deleted posts act as evidence of the same UIN and, therefore, can be used in the estimation of privacy risks.

The goal of this step is to identify if the regrettable scenario elicited in Steps 1 and 2 of the method corresponds to a pre-existing $PH_i$ inside the PHDB. Overall, there are three ways in which a $PH_i$ can match the elicited scenario. The first one is when the SAs, Audience and UIN associated with $PH_i$ are \textit{equal} to the SAs, Audience and UIN elicited in Steps 1 and 2 of the method (i.e. $PH_i.SAs=post.SAs$, $PH_i.Audience=post.Audience$ and $PH_i.UIN=post.UIN$). For instance, a user reporting \textit{job loss} after posting \textit{``My job at this company is like the coffee they serve...awful!''} is a scenario very similar to the one of Fig. \ref{fig:fig2}. Basically, in both cases the same UIN (i.e. \textit{job loss}) takes place after disclosing the same set of SAs (i.e. \textit{work location}, \textit{employment status} and a \textit{negative sentiment}) to the same Audience (i.e. \textit{work colleagues}). If we analyse a regrettable scenario in terms of \textit{sufficient} and \textit{necessary} conditions, one can say that disclosing a set of SAs to a specific Audience is sufficient condition for an UIN to (eventually) occur (i.e. $SAs,Audience \Rightarrow UIN$). In this sense, both scenarios have the same sufficient and necessary conditions. Consequently, if the sufficient and necessary conditions of $PH_i$ are equal to the ones of the elicited scenario, then $PH_i$ matches the elicited scenario.

The second matching case occurs when the UIN reported by the user differs from the one of $PH_i$, but the Audience and SAs stay equal (i.e. $PH_i.SAs=post.SAs$, $PH_i.Audience=post.Audience$ and $PH_i.UIN \neq post.UIN$). Let us consider again the previous example in which the user compares her workplace with the quality of her coffee. Let's imagine for a moment that this time the user has indicated \textit{harassment} as the UIN instead of \textit{job loss}. According to the matching criterion just introduced, the $PH_i$ which models the scenario of Fig. \ref{fig:fig2} does not match this new scenario. However, both scenarios describe the same \textit{sufficient conditions} (i.e. SAs and Audience) under which these UINs may take place. Hence, the new scenario can be modelled just by adding a new UIN (i.e. \textit{harassment}) to $PH_i$ (as we show in Step 5). Therefore, if only the sufficient conditions of $PH_i$ are equal to the ones of the elicited scenario, then $PH_i$ matches the elicited scenario after adding the elicited UIN to it.

The third matching case occurs when the SAs extracted in Step 1 of the method are a \textit{superset} of the ones of $PH_i$ but the Audience and the UIN reported by the user stay equal (i.e. $PH_i.SAs \subset post.SAs$, $PH_i.Audience=post.Audience$ and $PH_i.UIN=post.UIN$). For instance, a user reporting \textit{job loss} after posting \textit{``Moving to Boston was definitely not a good idea...the weather in this town sucks and so my job at this company! \#wrongdecisions''} is also a scenario a very similar to the one of Fig. \ref{fig:fig2}. In both cases the Audience is \textit{work colleagues} and the UIN is \textit{job loss}; however, an additional SA is disclosed in this new scenario: \textit{home location}. According to the matching criteria introduced so far, the $PH_i$ which models the scenario of Fig. \ref{fig:fig2} does not match the new scenario since their sufficient conditions are not equal. Nevertheless, the SAs disclosed in the new scenario are a \textit{superset} of the ones disclosed in Fig. \ref{fig:fig2}. This means that, according to $PH_i$, \textit{job loss} can already occur when revealing \textit{fewer} SAs to an audience composed by \textit{work colleagues}. In this case, we say that the sufficient conditions of $PH_i$ \textit{absorb} the ones of the elicited scenario. Therefore, if the sufficient conditions of $PH_i$ absorb the ones of the elicited scenario, then $PH_i$ matches the elicited scenario when the necessary conditions of both stay equal. 

If there is a PH that follows any of the matching criteria previously described, then it is retrieved from the PHDB and handed to Step 5. Otherwise, a new PH which represents the regrettable scenario being reported by the user must be created and added to the PHDB. This task, which corresponds to the \textbf{Step 4} of the method, takes care of generating the corresponding $PH_i$, taking as input the SAs of the deleted post together with the Audience and UIN entered by the user. The result in this case is a new PH and, consequently, an updated version of the PHDB which corresponds to PHDB* in Fig. \ref{fig:fig4}.

\textbf{Step 5: Update Contingency Table} Regrettable self-disclosure scenarios can result in more than one UIN. For instance, the scenario of Fig. \ref{fig:fig2} may result sometimes in \textit{job loss} and in other cases in \textit{reputation damage}, depending on the situation reported by each user. Likewise, an UIN can be the consequence of different regrettable scenarios. For instance, disclosing ones' sexual orientation or religious beliefs can lead in both cases to \textit{unjustified discrimination}. Therefore, a PH can be associated with different UINs, and different UINs can be associated with more than one PH. Furthermore, some UINs are likely to occur more often than others. This means that, for the PH associated with the scenario of Fig. \ref{fig:fig2}, \textit{reputation damage} can be reported by the users more frequently than \textit{job-loss} (or vice versa). Moreover, certain consequence values of a particular UIN can be reported more often than others. For instance, \textit{reputation damage} can be perceived in most cases as an event of a \textit{minor} magnitude and rarely as a \textit{catastrophic} event. Therefore, some consequence values may have a higher frequency than others.

A Contingency Table (CT) is a structure which organizes the information about the frequency of the UINs associated with a PH. Basically, it is a double-entry table in which each cell describes the number of times that an $UIN_j$ has been reported as the consequence of the scenario modelled by a $PH_i$. This is expressed through a tuple of five elements representing the frequency of the values \textit{catastrophic}, \textit{major}, \textit{moderate}, \textit{minor} and \textit{insignificant}. For instance, according to Table \ref{table:2}, the $UIN_2$ has been reported 108 times as a negative consequence of the scenario modelled by $PH_1$. From these 108 occurrences, 50 were reported as events of \textit{catastrophic} magnitude, 48 as \textit{major}, and 10 as \textit{moderate}. The goal of this step is to update the CT with the information provided by the user in Step 2. Basically, this consists of incrementing by 1 the consequence value of the UIN reported by the user for the PHs identified in Step 3 (or the PH created in Step 4). For instance, let us assume that one of the PHs identified was $PH_1$ and the user has reported $UIN_2$ as the UIN with a \textit{moderate} consequence level. Then, the output of this step is an updated version of the CT in which the tuple that corresponds to the $PH_1$ and $UIN_2$ contains now the values \{50,48,11,0,0\}. In the case that a new PH has been created as result of Step 4 (or the user has specified a new UIN in Step 2), a new row (or column) of \{0,0,0,0,0\} corresponding to such PH (or UIN) must be added to the CT prior to the incrementation of the UIN's frequency.

\begin{table}[!h]
\def\arraystretch{1}
\resizebox{\linewidth}{!}{
\begin{tabular}{|C{0.8cm}|C{1.7cm}|C{1.7cm}|C{1.7cm}|}
\hline
&
\multicolumn{1}{C{1.7cm}|}{\bm{$UIN_1$}} & 
\multicolumn{1}{C{1.7cm}|}{\bm{$UIN_2$}} & 
\multicolumn{1}{C{1.7cm}|}{\bm{$UIN_3$}}\tabularnewline \hline
\bm{$PH_1$} & \{0,0,0,0,0\} & \{50,48,10,0,0\} & \{0,0,44,188,90\} \tabularnewline 
\hline
\bm{$PH_2$} & \{0,0,0,0,0\} & \{0,0,79,55,0\} & \{0,0,0,0,0\} \tabularnewline 
\hline
\bm{$PH_3$} & \{0,0,0,0,0\} & \{0,0,0,0,0\} & \{120,88,7,0,0\} \tabularnewline 
\hline
\bm{$PH_4$} & \{300,33,0,0,0\} & \{0,0,0,0,0\} & \{0,0,0,0,0\} \tabularnewline 
\hline
\bm{$PH_5$} & \{0,0,0,0,0\} & \{0,310,70,0,0\} & \{0,0,0,0,0\} \tabularnewline 
\hline
\end{tabular}
}
%\bigskip
\caption{Contingency Table}
\label{table:2}
\vspace{-4mm}
\end{table}

\section{Adaptive Awareness Generation} \label{algorithm}

As we mentioned previously, PHs represent knowledge on recurrent self-disclosure scenarios that often lead to regrettable experiences. Therefore, they can be used to (i) detect potentially regrettable scenarios, and (ii) alert on the risks associated with such scenarios. The method we have just described is an instrument for collecting and organizing evidence on UINs which are associated with different PHs. Therefore, the content inside the CT can help us to estimate and communicate the severity of the different risks that may arise when disclosing certain patterns of private information in SNSs. In this section we describe how the information contained inside the CT can be applied to the generation of adaptive privacy warnings. Concretely, we introduce an algorithm which computes the severity of privacy risks associated with the publication of a post and inform the user about such risks. In order to regulate the frequency of such interventions, the algorithm incorporates a mechanism based on the action taken by the user after the warning is triggered (i.e. if the user publishes or not the post in the end).

\subsection{Identify Unwanted Incidents}

Algorithm \ref{alg:1} describes a process in which the information inside the CT is used to communicate the risks associated with a self-disclosure act. Basically, this consists of the generation of a warning message $wMSG$ describing the risks that may occur if the user posts a message with private information. For this, the function \textit{GenerateWarningMSG} is invoked when she attempts to share a post $P$ in a SNS. First, function $GetSAs$ (line 3) analyses the information disclosed inside $P$ and extracts the SAs from it (i.e. like in \textit{Step 1} of the method introduced in section \ref{method}). The result (i.e. a set of SAs) is assigned to $postSAs$ and used thereafter to compute a set of PHs which can provide information about potential privacy risks. This basically consists of collecting those PHs whose SAs are included in the ones of the post. In other words, $PH_i$ is retrieved from the $PHDB$ when $PH_{i}.SAs \subseteq postSAs$. This step is performed by the function $GetPHs$ and its result then assigned to $postPHs$ (line 4). If $postPHs \neq \emptyset$, it means that there is evidence about privacy risks that might occur after the publication of $P$. That being the case, the next step is to estimate the severity of such risks and communicate them to the user.

\begin{algorithm}[!h]
\caption{Adaptive awareness pseudo-code} \label{alg:1}
  \begin{algorithmic} [1]
  \Function{GenerateWarningMSG}{$Post~P$}%\hspace{\algorithmicindent}$%\Comment{This is a comment}
  \State $WarningMSG~wMSG;$
  \State $Set$\textless$SA$\textgreater$~postSAs := GetSAs(P);$
  \State $Set$\textless$PH$\textgreater$~postPHs := GetPHs(postSAs,~PHDB);$
  \ForAll {$PH_i~\in~postPHs$}
  	\State$Audience~Au := GetAudience(PH_i);$
  	\State $Set$\textless$UIN$\textgreater$~postUINs := GetUINs(ContTbl,~PH_i);$
  	\ForAll {$UIN_j~\in~postUINs$}
  		\State $ConsFreq~F := GetConsFreq(ContTbl,PH_i,UIN_j);$
  		\State $CritIndex~\hat{I}_{ij} := ComputeCritIndex(F);$
  		\If {$\hat{I}_{ij} > \varphi$}
  			\State $wMSG.addRisk(UIN_j,~Au);$
  			%\State $wMSG.addMainRsk(UI_j,~Au);$
  		%\Else
  	  		%\State $wMSG.addMinorRsk(UI_j,~Au);$
  	\EndIf
  	\EndFor
  \EndFor
  \State $RaiseWarning(wMSG);$
  \State $Action~usrAction := WaitForUsrAction();$
  \State $UpdateRiskThreshold(usrAction);$
  %\State \textbf{return} 
  \EndFunction  
  
  \end{algorithmic} 
\end{algorithm}

In order to estimate the privacy risks of the post, we must first determine the frequency of those UINs that can take place if the post is shared. For this, we iterate through each $PH_i$ inside $postPHs$ (line 5) and extract (i) its audience, and (ii) a list of the UINs that are associated with it (lines 6 and 7). The information about the audience is extracted by the function $GetAudience$, assigned to the variable $Au$ and used later for the generation of the warning message (line 6). On the other hand, the function $GetUINs$ queries $ContTbl$ (i.e. the CT) to gather those UINs associated with the $PH_i$ whose frequency is greater than zero. Its outcome (i.e. a set of UINs) is assigned to $postUINs$ and used thereafter for estimating the privacy risks of the posts (line 7). Basically, the estimation of the risks consists of computing the CI of those UINs that are associated with any of the PHs inside $postPHs$. For this, we must first iterate through each $UIN_j$ in $postUINs$ (line 8) and obtain the frequency of its consequence values from the CT (line 9). This is done by the function $GetConsFreq$ which retrieves from the CT the cell corresponding to the $PH_i$ and the $UIN_j$. After its execution, its outcome is assigned to $F$ and handed to the next step which is the computation of the CI.

\subsection{Compute Criticality Index} \label{compute_ci}

The function $ComputeCritIndex$ takes the frequency $F$ of $UIN_j$ and estimates its risk severity (line 10). For this, it uses the approach described in section \ref{critIndex} which consists of computing a CI using Eq. 1. To illustrate this step, let us assume that the user writes a post similar to the one in Fig. \ref{fig:1} and the SAs disclosed inside the post matches the ones of $PH_1$. Let us also assume that the information inside the CT is the one illustrated in Table \ref{table:2}. Then, we have two UINs (i.e. $UIN_2$ for job loss and $UIN_3$ for reputation damage) that are likely to occur if the post is shared by the user. In order to estimate the risks of this post we apply the Eq. \ref{alg:1} using the frequency values corresponding to $UIN_2$ and $UIN_3$ for $PH_1$:

\begin{align*}
\hat{I}_{12}=\frac{(\frac{25}{54}+\frac{49}{54}+1+1+1)-1}{5-1}= 0.843 \\[1ex]
\hat{I}_{13}=\frac{(0+0+\frac{22}{161}+\frac{232}{161}+1)-1}{5-1}= 0.214
\end{align*}

The values obtained for $\hat{I}_{12}$ and $\hat{I}_{13}$ correspond to the CI of $PH_1$ for $UIN_2$ and $UIN_3$, respectively. According to these values, the severity of \textit{job loss} is higher than the one of \textit{reputation damage}. However, as mentioned in section \ref{critIndex}, these values are an estimation of the CI based on a sample. Consequently, we must build a confidence interval containing the \textit{real} parameter $I$ with a certain confidence level. For this, we must first estimate the variance of $\hat{I}$ according to the following equation:

\begin{align*}
Var(\hat{I})&=\frac{1}{n(K-1)} \bigg[\sum\nolimits_{k=1}^{K-1}(K-k)^2 p_k (1-p_k) \\[1ex] &-2\sum\nolimits_{k=1}^{K-1}(K-k)p_k \sum\nolimits_{l=1}^{k-1} (K-l)p_l \bigg] \tag{2}
\end{align*}

where $n$ is the size of the sample, $K$ the number of consequence levels, and $p_k$ the proportion of observations in the sample corresponding to the category $k$. Using this equation, a confidence interval for $\hat{I}$ can be obtained as:

\begin{align*}
\hat{I}-Z_{\alpha/2} \cdot S(\hat{I}) \leq I \leq \hat{I}+Z_{\alpha/2} \cdot S(\hat{I}) \tag{3}
\end{align*}

where $S(\hat{I})$ is the standard deviation of $\hat{I}$, $\alpha$ the significance level, and $Z_{\alpha/2}$ the standard score for $\alpha/2$. Consequently, for a significance level $\alpha=0.05$, the confidence intervals for $I_{12}$ and $I_{13}$ are $[0.782;0.904]$ and $[0.180;0.249]$, respectively. Following a conservative criterion, the outcome of the function \textit{ComputeCritIndex} will be the upper bound of the confidence interval created for each UIN. Therefore, the function will return 0.904 in the case of $UIN_2$, and 0.249 for $UIN_3$.

\subsection{Communicate Unwanted Incidents}

As mentioned in section \ref{background}, not all users are equally concerned about their privacy. There are users who are willing to expose themselves more in SNSs and users who rather keep their private information away from public disclosure. In other words, some users take higher privacy risks than others when making privacy decisions. In line with this premise, Algorithm \ref{alg:1} introduces a privacy risk threshold $\varphi$ which is used to determine which UINs should (or should not) be communicated to the user. Basically, it consists of a value between 0 and 1 which is tested against the risk criticality index returned by \textit{ComputeCritIndex}. If this value is equal or higher than $\varphi$, then it means that the risk is unacceptable for the user and, therefore, the corresponding UIN should be communicated. Conversely, if this value is lower than $\varphi$, then the risk is acceptable, and the UIN should not be informed (line 11). To illustrate this mechanism, let us assume that $\varphi=0.5$ and the criticality indexes being evaluated correspond to the ones of $\hat{I}_{12}$ and $\hat{I}_{13}$. Since $\hat{I}_{12}=0.904 > \varphi$, then $UIN_2$ is added to the body of the warning message $wMSG$ together with its corresponding audience $Au$ (line 12). On the other hand, the risk criticality index $\hat{I}_{13}=0.249 < \varphi$, hence, $UIN_{3}$ is not included in the body of the warning message. %Those UINs to be communicated are attached to the content of a warning message $wMSG$ together with the audience of their corresponding PH.

\begin{figure}[!h]
\centering
\includegraphics[height=3.5cm]{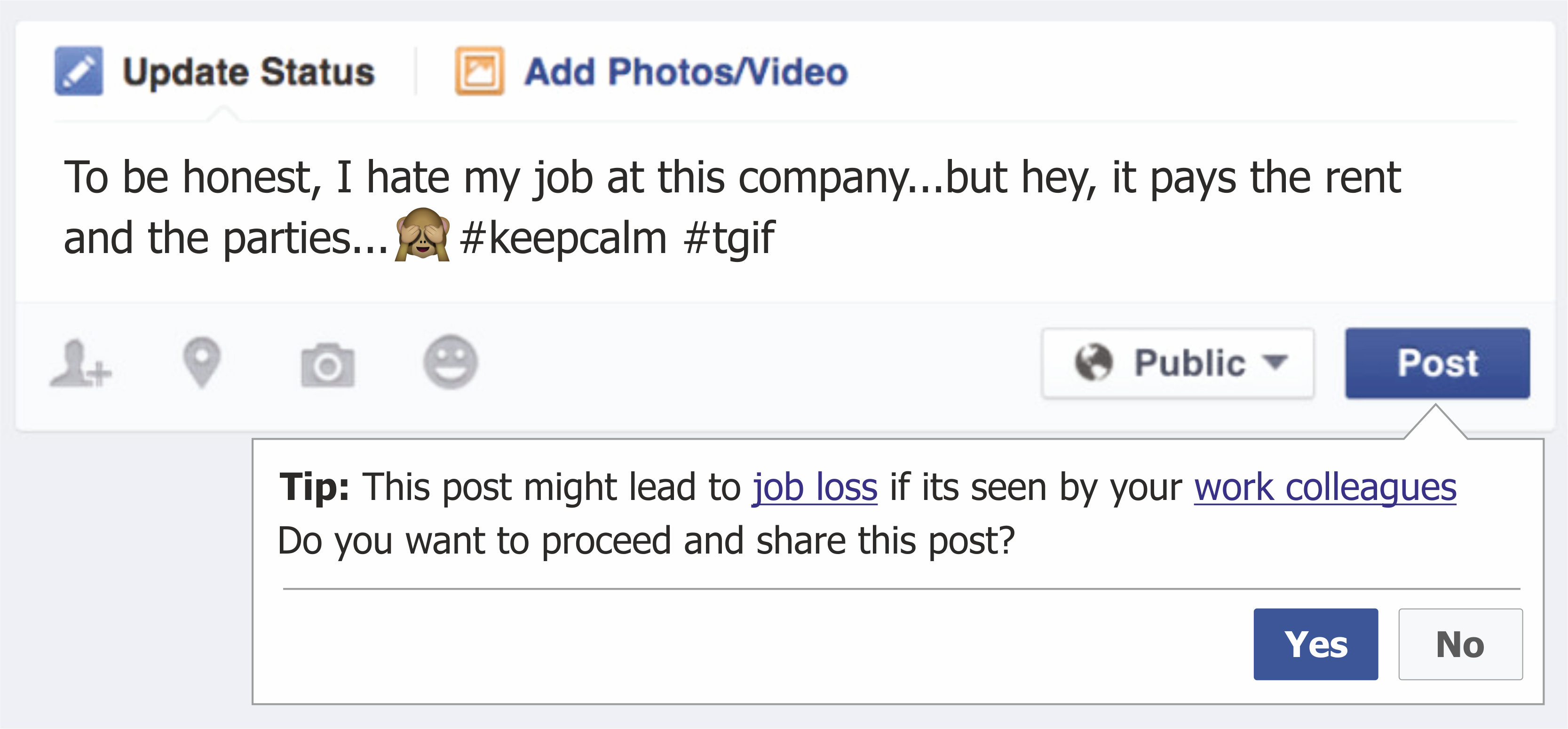}
\caption{Envisioned Interface}
\label{fig:5}
\end{figure}

After the content of the warning message has been defined, the function $RaiseWarning$ takes over the task of communicating the message to the user (line 16). This can be done using a pop-up message like the one illustrated in Fig. \ref{fig:5}. At this point, the user has the chance to re-think the content of her post or proceed with its publication. It may happen that after some time the number of \textit{rejected warnings} is higher/lower than the number of \textit{accepted warnings}. This is, one may observe that, after a time frame $\tau$, the user tends to ignore (or accept) the warnings and proceed (or not proceed) with the publication of her posts. In order to regulate the frequency of the interventions, the value of $\varphi$ is adjusted at the end of each $\tau$ interval according to the actions taken by the user. Basically, this consists of decreasing/increasing the value of $\varphi$ depending on the number of times the user has accepted/rejected an intervention. For this, the function $WaitForUsrAction$ waits for the user's decision and forwards it to the function $UpdateRiskThreshold$ which takes care of updating the value of $\varphi$ (lines 17 and 18). This function keeps track of the number of times the user has ignored/followed the warnings within a $\tau$ period of time. After each $\tau$ period, if $\#ignored > \#accept$, then the value of $\varphi$ is increased in $\delta$ (i.e $\varphi_{\tau + 1}:= \varphi_{\tau} + \delta$). Conversely, if $\#ignored < \#accept$, then $\varphi$ is decreased in $\delta$ (i.e $\varphi_{\tau + 1}:= \varphi_{\tau} - \delta$).

\section{\uppercase{Discussion}} \label{discussion}

Although our approach is devoid of assumptions related to the estimation of privacy risks, there are limitations that should be acknowledged and considered. One of these issues is related to the identification of SAs inside the users' post. As we mentioned in section \ref{method}, there are different NLP methods that could be applied for the automatic identification of such SAs. However, the use of sarcasm or irony (which is common inside SNSs) can modify significantly the meaning of a post and, therefore, hinder its analysis (e.g. a post could be classified as negative, when in fact it is sarcastic) \cite{joshi2017automatic,joshi2016challenging}. Another issue is related to posts which are not self-referential. For instance, a post like \textit{``Working at Google may sound great...but I am sure that it can be a very competitive and hostile work environment''} is expressing a negative opinion about working for Google, however, it is not saying that the user who wrote it works for this company. Therefore, whatever method one defines for the identification of SAs inside posts, it should be aware of these variations in order to identify regrettable self-disclosure scenarios in a correct way.

Another aspect to be considered is related to the values that are assigned to the parameters of Algorithm \ref{alg:1}. For instance, one must assign an initial value to $\varphi$ which can result in more or less interventions at the initial phases of the awareness process. For instance, a value of $\varphi$ closer to 0 would result in a higher intervention frequency, whereas a value closer to 1 would generate a lower amount of interventions. In line with this, the value assigned to $\tau$ can impact the values adopted by $\varphi$ at the adaptation phase. This is, a small $\tau$ would limit the amount of evidence gathered on warning being accepted/ignored by the user. Hence, the value of $\varphi$ would probably not reflect the user's privacy behaviour. Likewise, a big $\tau$ would result in values of $\varphi$ that stay invariant for long periods of time. Consequently, the frequency of the interventions would not be reactive enough with regard to the user's privacy decisions. Both parameters, $\varphi$ and $\tau$, should be chosen with care in order to guarantee a sustained awareness support to the user. %Moreover, a criterion to reset the values of these parameters after a certain time could be useful for achieving this goal.

Although the awareness system described in this paper stands for the protection of the users' privacy, it is ultimately a recommender system. Moreover, it is a system which requires analysing and processing personal information for shaping its recommendations. Hence, its benefits come along with the privacy concerns that are characteristic of recommender systems. That is, issues related to algorithmic transparency, fairness and trust that can jeopardize the users' privacy rights. This calls for a Data Privacy Impact Assessment (DPIA) of the different software artefacts and information flows described throughout this paper (i.e. PHDB, UPDB, post analysis, etc.). The notion of DPIA has been introduced in the EU General Data Protection Regulation \cite{regulation2016regulation} and is basically an assessment that service providers must conduct in order to identify and minimize the risks that data processing may bring to the privacy rights of data subjects (i.e. the users). Although this analysis goes beyond the scope of this work, it is a critical point that must be taken into consideration and further elaborated.

\section{Design and Evaluation Plan} \label{evaluation}

Evaluating the approach introduced in this paper brings up a series of challenges related to the technical and human resources that are necessary for setting up an experimental environment. Basically, our approach requires input from a large number of users in order to estimate the risk value of a set of self-disclosure scenarios. Moreover, enough evidence on each particular scenario inside the PHDB is necessary to perform such estimation. Hence, one must count not only on a large number of participants, but these participants should also provide enough input on each of the PHs stored in the PHDB. In principle, gathering such extent of user input could be possible through the development of some SNS plugin that implements the interface of Fig.~\ref{fig:3}. However, it would still take some time until enough information on each PH is collected and, consequently, until the corresponding risk index could be computed.

One way of gathering the information necessary for the estimation of privacy risks is through an online questionnaire. For instance, one can propose a questionnaire consisting of a set of self-disclosure scenarios and ask the participants to rate the severity level of each of them using an ordinal scale. Each scenario can be afterwards represented as a PH, and the information collected from the questionnaires inserted in the CT. By doing this, we simplify the process of collecting heuristic-related information through deleted posts and simulate the knowledge extraction process described in section \ref{background}. On the other hand, the process of knowledge application can be carried forward by implementing the warning system described in this paper as a mobile application. That is, an application from which (i) users can post messages using their SNS accounts (e.g. via the Facebook or Twitter APIs), and (ii) intervenes following the approach described in section \ref{algorithm}. This app can be used afterwards in an experiment in which the participants are requested to install it and use it for a certain time. Thereafter, the effectiveness of the interventions generated by the app can be evaluated conducting structured interviews with the participants.

\section{\uppercase{Conclusion and Future Work}} \label{conclusion}

\noindent Risk communication and management is a valuable instrument for helping the users of SNSs to make better and more informed privacy decisions. Particularly, incorporating adaptive risk awareness features into the design of PTs can have a positive impact on their engagement levels \cite{diaz2018iwse}. As we mentioned on section \ref{background}, engineering such adaptive solutions requires the definition of processes related to knowledge extraction and application. In this work we have addressed both activities through (i) the definition of a method for collecting information about UINs and (ii) an algorithm for generating adaptive interventions. The envisioned interface illustrated on Fig. \ref{fig:5} shows how these two instruments can work in cooperation in order to endow SNSs with user-centred privacy awareness features.

Adapting the risk information and frequency of interventions is a step towards more effective PTs. However, a study of Kaptein et al. \cite{kaptein2012adaptive} suggest that framing the \textit{style} of an intervention can also improve its effectiveness. In their study, persuasive messages for promoting healthier eating habits were framed using either an authoritarian-style (e.g. \textit{``The World Health Organization advises not to snack. Snacking is not good for you''}) or a consensus-style (e.g. \textit{``Everybody agrees: not snacking between meals helps you to stay healthy''}). The outcome of this experiment suggests that persuasive messages are more effective when tailored to the user's preferred persuasive style. This phenomenon was also observed in an experiment conducted by Sch\"{a}wel et al. \cite{schaewel2018disclose} in which warning messages were used to promote safer self-disclosure decisions among the users of SNSs. Particularly, an intervention framed using an authoritarian-style such as \textit{``Rethink what you are going to provide. Privacy researchers from Harvard University identify such information as highly sensitive!''} can be more effective than another one framed using a consensus-style like \textit{``Everybody agrees: Providing sensitive information can result in privacy risks!''}, and vice versa. Adapting the persuasive style of interventions is an aspect that will be further investigated in our future publications.

\begin{acks}
This work was supported by the Deutsche Forschungsgemeinschaft (DFG) under grant No. GRK 2167, Research Training Group ``User-Centred Social Media''.
\end{acks}

\bibliographystyle{ACM-Reference-Format}
\balance
\bibliography{references} 

\end{document}